\documentclass[prb,reprint,notitlepage,tightenlines]{revtex4-1} 
\usepackage{amsmath}  
\usepackage{amsfonts} 
\usepackage{graphicx} 
\begin{document}

\title{The Daniell Cell, Ohm's Law and the Emergence of the International System of Units} 

\author{Joel S. Jayson}
\email{jsjayson@yahoo.com} 
\affiliation{ Brooklyn, NY }

\date{\today}

\begin{abstract}
Telegraphy originated in the 1830s and 40s and flourished in the following decades, but with a patchwork of electrical standards.  Electromotive force was for the most part measured in units of the predominant Daniell cell. Each company had their own resistance standard.   In 1862 the British Association for the Advancement of Science formed a committee to address this situation.  By 1873 they had given definition to the electromagnetic system of units (emu) and defined the practical units of the ohm as ${10}^9$ emu units of resistance and the volt as ${10}^8$ emu units of electromotive force.  These recommendations were ratified and expanded upon in a series of international congresses held between 1881 and 1904.  A proposal by Giovanni Giorgi in 1901 took advantage of a coincidence between the conversion of the units of energy in the emu system (the erg) and in the practical system (the joule) in that the same conversion factor existed between the cgs based emu system and a theretofore undefined MKS system.  By introducing another unit, X (where X could be any of the practical electrical units), Giorgi demonstrated that a self consistent MKSX system was tenable without the need for multiplying factors.  Ultimately the ampere was selected as the fourth unit.  It took nearly 60 years, but in 1960 Giorgi's proposal was incorporated as the core of the newly inaugurated International System of Units (SI).  This article surveys the physics, physicists and events that contributed to those developments.
\end{abstract}
\maketitle

\section{Introduction} 

 The main thrust of this paper aims at demonstrating the decisive role that mid-19th century telegraphy had on the establishment of standard measurement units and further on the specific values those standards took, and how these developments led directly to the eventual inauguration of the International System of Units (SI).

At mid-century batteries of electrochemical cells held the predominant position as a source of electrical power.  Applications such as arc lighting and electric motors were either at a developmental stage and/or had to await the maturation of the electric generator later in the century before they emerged as major commercial entities. The electrochemical industry and telegraphy were the primary electrical based enterprises, and of these two, telegraphy, in particular, had developed into a major industry and was growing at a rapid rate.  One need only observe the exponential growth rate of the Internet in our age, and additionally recognize that telegraphy was the first of the ``instantaneous'' communication media, to realize the affect that telegraphy had at that time.

With regard to the evolution of modern measurement units, the place to begin is not mid-19th century, but rather towards the end of the 18th century. In pre-revolutionary France measurement standards varied not only from province to province, but from town to town.  The confluence of the French Revolution and the advocacy of universal standards by such eminent proponents as Pierre-Simon Laplace (1749--1827) and Adrien-Marie Legendre (1743--1833) made the times ripe for reform.  The meter and kilogram standards were established in 1799, with decimal multiples and submultiples explicit to the system.\cite{alder}

With regard to the development of the International System of units (SI), it might appear that the rest was history, but not without traveling a somewhat convoluted path, specifically, with respect to electrical units.  It is obligatory that contemporary texts on electrodynamics introduce tables and devote some discussion towards the conversion between SI units and the several alternate systems, a couple of which are still in common usage in areas of theoretical physics.\cite{jackson}

Absent from those contemporary texts is an accounting of why the SI units occupy a predominant position rather than any of the alternatives, all of which are metric and fall under the ambiguous umbrella term of ``cgs'' units.  In particular, the central role played by ``practical'' electrical units in reaching this circumstance goes unaddressed.  Here, we provide an outline of the path to the SI system of units, avoiding a detailed discussion of the alternative systems except where they enter directly into the study.   A number of scientists and engineers are introduced along the way.

The following section provides a brief background on the units of time, length, and mass.  That discussion is then followed by a segment on the origins of the electromagnetic (emu) and electrostatic (esu) systems of units and immediately after by a section on the practical units.  These strands provide the background necessary to appreciate the ensuing section on the foundation of the SI system of units. Finally, we briefly regard topics not covered in the main body of the paper, which are deemed relevant to the discussion.

\section{Time, Length and Mass}

The astronomical definition of the second (1/86,400 of a mean solar day) was the time standard into the 20th century.  Increasingly accurate measurements of time have prompted a progression from that astronomical definition of the second to the current definition of ``9,192,631,770 periods of the radiation corresponding to the transition between two hyperfine levels of the ground state of the cesium 133 atom.''\cite{bipm}  As of January, 2013 the relative uncertainty in the NIST-F1 fountain cesium clock, the  primary time and frequency standard of the United States, was $3\times10^{-16}$, or less than one second in a 100 million years.\cite{nist2}

The meter was originally defined as one ten-millionth the distance from the equator to the north pole and the gram as the mass of a cubic centimeter of water at the temperature at which water assumes its maximum density.  Both of these definitions are inconvenient for commercial use, so two prototypes were fabricated of platinum, the length of one representing the meter and the mass of the other defining the kilogram.  Extended surveys revealed that the original value of the meter was low, but the artifact was retained as the standard, an acknowledgment that a reliable standard is what really mattered (with a reference to a geographical length being arbitrary).

 By 1875, with advancement in precise measurement and material sciences, the question of length and mass standards was revisited on an international scale.  The Convention du M\`etre was signed in Paris by 17 nations.  This treaty established the General Conference on Weights and Measures (known by its French initials as CGPM) as the body that ratifies standards proposals.  The  CGPM meets every few years. The first conference was held in 1889, the most recent (the 25th) in 2014.   Also established under the Meter Convention, and serving under the CGPM, is the International Committee for Weights and Measures (CIPM).  The CIPM, aided by a number of consultative committees, effects the scientific decisions.  Finally, the International Bureau of Weights and Measures (BIPM) was established under the supervision of the CIPM  to coordinate the activities of the national standards laboratories.   New platinum-iridium artifacts for the meter and kilogram were fabricated and sanctioned by the first CGPM in 1889.\cite{bipm}

The 1889 kilogram prototype continues as the mass standard today. The prototype artifact and copies are kept in a vault by BIPM and national prototypes have been distributed to member countries of the Meter Convention.  Comparisons are periodically conducted and since 1889 a drift in relative mass of about $5\times10^{-8}$ has been noted.\cite{davis}  The CGPM has recommended that the mass standard be redefined in terms of fundamental constants and several national laboratories have undertaken serious efforts in this direction, with a goal of a relative uncertainty of less than ${10}^{-8}$.  As it stands, it appears that this goal will be attained and that the kilogram standard eventually will be redefined.\cite{mills,robinson,quinn}

In contrast to the kilogram, the definition of the meter has undergone sweeping changes.  By 1983 both wavelength and frequency could be measured with sufficient accuracy to enable the speed of light in vacuum to be fixed as: $c=299,792,458~\mathrm{m/s}$.  The meter since has been defined as the length that light travels in vacuum in 1/299,792,458 seconds.\cite{bipm}

This brief description on the time, length and mass units suffices for our purposes.  For a more in-depth understanding of the history and practices of metrology see Refs.~\onlinecite{bipm}, ~\onlinecite{nist2} and~\onlinecite{quinn}.

\section{The EMU and ESU Systems of Units}

The renowned naturalist Alexander von Humboldt (1769--1859)  undertook as one of his multiple projects the mapping of the Earth's magnetic field.  He enlisted the great mathematician Carl Friedrich Gauss (1777--1855) with the task of establishing accurate measurements. In the 18th century Charles-Augustin de Coulomb (1736--1806) had reported on the inverse square distance relationship for both electric charges and magnetic poles. 

In 1832 Gauss presented a paper on the measurement of the terrestrial magnetic field in which, using metric measures, he introduced absolute measurements.\cite{gauss} By metric measures we mean units that are some decimal multiple of the meter and the kilogram.  Gauss employed millimeters and milligrams.  He used the Coulomb formula for the force between two magnetic poles,
\begin{equation}\label{gausseq}F=\frac{k_1 p_1 p_2}{r^2}~,\end{equation}
where $k_1$ is a constant, $p_1$ and $p_2$ the two monopole strengths and $r$ the distance between the two poles. Here Gauss set $k_1$ as a dimensionless quantity equal to unity, which led to the pole strengths being measured in terms of units of mass, length and time.  As was recognized at the time, monopoles are an idealization.  Gauss's accurate analysis and measurements took into account the dipole nature of the magnetic sources.\cite{feather}    Since that presentation by Gauss, metric measure has been employed in the scientific literature.   

Earlier in the century, in 1819, Hans Christian Oersted (1777--1851) had observed the deflection of a magnetic needle due to the influence of an electric current.  His observation was immediately followed by experiments and analysis by Jean-Baptiste Biot (1774--1862) and Felix Savart (1791--1841) in 1820 that provided a relationship between an electric current and the magnetic flux density that it generated.  Andre-Marie Amp\`ere (1775--1836) conducted extensive experiments from 1820--1825 on the forces between currents.\cite{whit}

In the Biot-Savart formulation the generated magnetic flux density at a point in space is expressed in terms of a current in an elemental length of conductor and its distance and orientation with respect to the given point.   Amp\`ere performed his experiments and analysis strictly in terms of currents, as field quantities did not enter into his equations.\cite{feather2}   For our purposes, we use the integrated elemental equation that expresses the force per unit length between currents in two parallel wires of negligible  cross-section and of infinite length separated by a distance d as
\begin{equation}\label{ampeq}\frac{dF_m}{d\ell}=\frac{2 i_1 i_2}{d}\ \quad~ (\mathrm{emu} ~ \mathrm{ units}). \end{equation}
Here the idealized magnetic monopoles of Eq.~(\ref{gausseq}) have been replaced by the currents that generate the magnetic fields.  The factor of 2 is an integration constant\cite{golding} and emu symbolizes electromagnetic units. At this point we are using emu units in a generic fashion.  As will be seen, modern emu units are expressed in units of centimeters, grams and seconds, as are esu units. 

In a manner similar to the definition of absolute emu units, absolute electrostatic units (esu) were defined through application of Coulomb's Law on charge,
\begin{equation}\label{coul}F=\frac{k_2 q_{1} q_{2}}{r^2},\end{equation}
with $k_2$ equal to a dimensionless quantity equal to unity.  Here the electric charges are measured in terms of mass, length and time.  There were now two equations expressed in absolute units, one in terms of charge flow and one in terms of charge.  It was imperative that the relationship between them be uncovered.

In 1856 Wilhelm Eduard Weber (1804--1891) and Rudolph Kohlrausch (1809--1858) performed a key experiment that evaluated this relationship.  They measured a charge in esu units and then discharged it through an impulse galvanometer, which was calibrated in emu units.  In so doing they obtained the factor of $c^2$, $c$ being the velocity of light, that related the electric force and the magnetic force in both the esu and emu sets of units.\cite{whit2}  Thus, for esu,
\begin{equation}\label{esu}F_{e1}=\frac{q_{e1} q_{e2}}{r^2}\end{equation}
and
\begin{equation}\label{esub}\frac{dF_{e2}}{d\ell}=\frac{2 i_{e1} i_{e2}}{c^2 d}\end{equation}
while for emu,
\begin{equation}\label{emu}F_{m1}=\frac{c^2 q_{m1} q_{m2}}{r^2}\end{equation}
and
\begin{equation}\label{emub}\frac{dF_{m2}}{d\ell}=\frac{2 i_{m1} i_{m2}}{d}.\end{equation}
This experiment was influential in leading James Clerk Maxwell (1831-1879) to later conclude that light was an electromagnetic wave.  Maxwell verified the result through an equivalent experiment.\cite{max}  (The same conclusion follows directly from special relativity).\cite{french}

\section{Practical Units}

\subsection{The Daniell cell}

Alessandro Volta (1745--1827) sometime in the 1790s, devised the electric pile (i.e., battery), which he made public in a letter to the Royal Society of London in  March, 1800.\cite{bord}  Volta's discovery enlivened early nineteenth century physics.  The initial stimulus was especially pronounced in chemistry.  The electrolysis of water was soon followed by the dissociation of a variety of compounds, and chemists isolated elements such as sodium and potassium for the first time.  Studies of static electricity in the 18th century were limited by the short discharge times of the accumulated charge.  The electric cell made it possible to explore the effects of steady currents. As discussed in the previous section, by the 1820s this led to fundamental discoveries.

The basic zinc and copper electrodes of Volta remained essentially unchanged until the 1830s.  A major problem in those early cells was polarization--the formation of hydrogen gas bubbles at the anode--that resulted in increased internal resistance and decreased efficiency.  In 1836 John Fredric Daniell (1790--1845) invented the cell which bears his name.  The cell still used zinc and copper electrodes, but made use of two electrolytes, copper sulfate and zinc sulfate.  The copper electrode was immersed in the copper sulfate and the zinc electrode in the zinc sulfate. The two electrolytes were initially segregated by an unglazed ceramic pot, but at a later time a gravity cell was devised that took advantage of the difference in specific gravity of the two sulfates.  The Daniell cell successfully resolved the polarization problem.  It had an open circuit potential of about 1.07 volts (in SI units).\cite{meyer}

\subsection{Ohm's law}

Georg Simon Ohm (1789--1854) discovered his renowned relationship during the mid 1820s.  Through a series of careful experiments Ohm established the familiar formula, $i=V\sigma A/{\ell}$,
 where $i$ is the current, $V$ the voltage, $\sigma$ the conductivity, $A$ the cross-sectional area, and $\ell$ the length of material across which the voltage is applied. 

\subsection{Telegraphy}

A nineteenth century telegraph circuit was simplicity itself, a battery at either end of a single conductor, the earth acting as a return path, and an electromagnetic key at each end to provide signaling.  Relay stations, lightning arrestors, and circuitry to allow simultaneous transmission in both directions, provided added sophistication, but of primary interest for this discussion is the battery and the conductor.

The Daniell cell was not the only stable source available, but because it was safe and relatively inexpensive, it became dominant.  Depending upon the circuit length, a varying number of cells were strung together to provide sufficient potential difference.  Iron was generally used as the conductor to provide strength.

Telegraph operators became proficient at knowing what combination of Daniell cells should be used with a given length of conductor.  Ground faults could be located if the line resistance per unit length was a known.  Each telegraph company specified a standard length of conductor at a given gauge as a resistance ``standard''.  Through the 1850s there were no industry wide standards.\cite{pope}

\subsection{British Association for the Advancement of Science}

Josiah Latimer Clark (1822--1898) was an accomplished engineer who worked in British telegraphy.  In 1861 Clark and another distinguished engineer, Charles Tilston Bright (1832--1888) read a paper before the British Association for the Advancement of Science (B.A.A.S.) recommending the establishment of a comprehensive system of electrical units.   Eminent members of the B.A.A.S. at the time included William Thomson (Lord Kelvin (1824--1907)), James Clerk Maxwell and James Prescott Joule (1818--1889).  Thomson was perhaps best known for his development of the absolute scale of temperature, but he also had played a central role in the laying of the trans-Atlantic telegraph cables.  After the Clark-Bright presentation Thomson took the initiative in forming a ``Committee on Electrical Standards''.  

The committee, which was active from 1862 to 1869, took the rudimentary emu and esu systems advanced by Weber as a starting point.   Because the emu system was convenient for specifying current, they stipulated that the practical units would be defined in decimal multiples of emu units.  They chose the meter, gram and second as the fundamental units,\cite{fleem}  The primary concern of the committee lay in defining a unit of resistance and fabricating physical standards that came close to realizing this unit.  They corresponded with scientists in several countries and by 1865 they had fabricated a number of standard resistors and had distributed them internationally to scientists and to directors of public telegraphs.  

Although the committee focused on defining resistance, it was clear from the beginning that the practical potential unit would be defined by a decimal multiple of the emu system unit that came closest to the electromotive force of the Daniell cell. Between this unit and the resistance unit the current unit would also be defined.  This was all made explicit in the short, but significant report of a second B.A.A.S. committee.  This ``Committee for the Selection and Nomenclature of Dynamical and Electrical Units'' redefined the fundamental units to be the centimeter, gram and second, thus displacing the meter.\cite{baas}   The committee also defined the cgs unit of force as the ``dyne'' and the cgs unit of work as the ``erg'', these dynamic quantities being the same in both emu and esu systems.  With mathematical physics without a complete system of standardized units for close to two centuries, the British Association for the Advancement of Science laid a solid foundation in little more than a decade.

\subsection{International Electrical Congresses, 1881--1904}

One would think that after the B.A.A.S. committees' formidable effort some semblance of order would have emerged.  Yet, in 1881 ``there were in various countries, 12 different units of electromotive force, 10 different units of electric current and 15 different units of resistance.''\cite{petley}  It happened that, also in 1881, the first International Electrical Congress was held in Paris.  This congress endorsed the several conclusions of the B.A.A.S. and additionally defined the practical unit of the ``ampere'' as one tenth the emu system unit for current. The nomenclature was original, but the value followed directly from the definitions of the volt and the ohm. 

Six congresses met in all from 1881 to 1904.  The second congress in 1889 agreed on  additional nomenclature for practical units including the ``joule'' (${10}^7$ ergs) and the ``watt'' (${10}^7$ erg/seconds). (Joule, who had demonstrated the equivalence between mechanical and heat energy, died only a few weeks after the conclusion of the 1889 congress.)  The electrical units in the emu and esu systems were given the same names as those in the practical system with the addition of prefixes, ``ab'' (for absolute) in the emu system and ``stat'' in the esu system.  Thus, volt, abvolt and statvolt were defined as the units for the electric potential difference in the practical, emu and esu systems, respectively.

By the late 1890s the units and standards agreed upon in these congresses held legal sway in the major industrial nations. 

For a comparison of emu and practical units refer to Table~\ref{conversion}.  Only two unit values were originally defined, the ohm and the volt.  As can be readily verified, all the other unit values follow from simple relationships and are explicitly labeled in the table as ``derived''.  For example, ${V}^2/R$, where $R$ is the resistance, has the dimensions of power. By direct substitution watts in the practical system converts to ${10}^7$erg/seconds in the emu system. 

\begin{table}[h!]
\centering
\caption{Conversion between Practical and EMU Systems of Units}
\begin{ruledtabular}
\begin{tabular}{ l l }
Practical Units & EMU Units \\
\hline	
Ohm & ${10}^9$ abohms \\
Volt & ${10}^8$ abvolts \\
Ampere (derived) & ${10}^{-1}$ abamperes \\
Coulomb (derived) & ${10}^{-1}$ abcoulombs \\
Farad (derived) & ${10}^{-9}$ abfarads \\
Henry (derived) & ${10}^9$ abhenries \\ 
Joule (derived) & ${10}^7$ ergs \\
Watt (derived) & ${10}^7$ erg/seconds \\
\end{tabular}
\end{ruledtabular}
\label{conversion}
\end{table}

The selection of practical values for the ohm and the volt was determined by the happenstance of common usage of a reliable electrochemical cell, the Daniell cell, and by the practical considerations of a telegraphy circuit; a resistance much less than an ohm, as defined, was not to be found due to the typical length of the circuit, and a resistance many times greater than an ohm required a large battery of cells to develop the required current (for a long circuit).  This latter was not unheard of, but with regard to the standards that the telegraph companies maintained, a shorter and more manageable length of wire was specified.

As we have shown, all practical units followed from the definition of the volt and the ohm. it then follows that all practical units ultimately originated from the electrochemical cell voltage and the wire resistance standards prevalent in the telegraphy industry.

\section{The International System of Units}

At the sixth, and final International Electrical Congress, held in St.\ Louis in 1904, a paper was presented ``On the Systems Of Electric Units'' by the Italian delegate, in which was appended a proposal by Giovanni Giorgi (1871--1950) concerning electrical and physical units.  Giorgi had previously published on the subject as early as 1901.\cite{giorgi}  Giorgi's proposal was based on the coincidence that if the cgs unit, the erg, were converted to an MKS unit, then that MKS unit would be equal to the practical unit of energy, the joule.  We are so familiar today with the joule as an ``MKS'' unit that this point bears repeating.  The joule was the nomenclature for the practical system energy unit and MKS units were not pertinent to its definition.

Given the fortuitous energy equality,  Giorgi went on and  converted from emu cgs units to MKS units in Eqs.~(\ref{emu}) and (\ref{emub}), but in doing so he didn't convert the $i^2$ and $q^2$ terms from emu to MKS, but rather from emu to practical units.  That would be inconsistent  unless he could add another degree of freedom, which he did by proposing a fourth fundamental unit.  Any of the practical electrical units would suffice. He left it for others to specify which ``X'' to use. Thus,
\begin{equation}\label{emu2}F_{m1}=\frac{c^2 q_{m1} q_{m2}}{r^2}\end{equation}
and
\begin{equation}\label{emu2b}\frac{dF_{m2}}{d\ell}=\frac{2 i_{m1} i_{m2}}{d},\end{equation}
in emu units convert to
\begin{equation}\label{MKSX}F_{M1}=\frac{{10}^{-7} c^2 q_{M1} q_{M2}}{r^2}\end{equation}
and
\begin{equation}\label{MKSXb}\frac{dF_{M2}}{d\ell}=\frac{2\times{10}^{-7} i_{M1} i_{M2}}{d},\end{equation}
in MKSX units. Note a factor of ${10}^{-7}$ had been introduced by this conversion.  

The inverse square equations naturally fit into spherical coordinates and expressing them with a factor of $1/{4 \pi}$ removes a factor of $4 \pi$ that appears in two of Maxwell's equations.  Oliver Heaviside (1850--1925) was the first to advocate this approach, which he termed ``rationalization''. 

Giorgi corresponded with Heaviside and subscribed to a rationalized system.\cite{giorgi2}   Since rationalization was ultimately incorporated into the SI system, we continue the discussion of Eqs.~(\ref{MKSX}) and (\ref{MKSXb} by assimilating the factor $1/{4\pi}$, as in
\begin{equation}\label{RMKSX}F_{M1}=\frac{4 \pi \times{10}^{-7} c^2 q_{M1} q_{M2}}{4 \pi r^2}\end{equation}
and
\begin{equation}\label{RMKSXb}\frac{dF_{M2}}{d\ell}=\frac{2(4 \pi \times {10}^{-7}) i_{M1} i_{M2}}{4 \pi d}\ \ ~  (\mathrm{MKSX} ~  \mathrm{units}).\end{equation}
Defining, $\epsilon_0={10}^7/(4 \pi c^2)=8.854\times{10}^{-{12}}$  farads/meter  and $\mu_0=4 \pi \times{10}^{-7}$  henries/meter, we have,  
\begin{equation}\label{SIMKS}F_{M1}=\frac{q_{M1} q_{M2}}{4 \pi \epsilon_0 r^2}\end{equation}
and
 \begin{equation}\label{SIMKSb}\frac{dF_{M2}}{d\ell}=\frac{ \mu_0 i_{M1} i_{M2}}{2 \pi d}\ \ ~  (\mathrm{SI} ~  \mathrm{units}),\end{equation}
where $\epsilon_0$ is the permittivity of free space and $\mu_0$ is the permeability of free space.  This is the form of the elementary force equations eventually incorporated into the SI system of units.

The International Electrotechnical Commission (IEC) is a permanent standards organization founded in 1906 that arose out of the international congresses.  In 1935 it adopted Giorgi's MKSX system, although the questions of both rationalization and the identification of the fourth unit were left open at that time.\cite{giorgi3} In 1950 it chose the ampere as the fourth basic unit. The final form of rationalization was adopted in 1956.\cite{curtis}

Following the recommendations of the IEC, the 9th CGPM in 1948 made the decision to establish the SI.  The 11th CGPM in 1960 confirmed the name Syst\`eme International d'Unit\'es, with the abbreviation SI and officially inaugurated the system.  It had been a long road from the French platinum meter and kilogram prototypes of 1799 and Volta's zinc-copper pile of 1800.\cite{quinn,petley,nelson,bord2,zim,sils} 

\section{Concluding Remarks}

A comprehensive history of the understanding of electromagnetism during the 19th century was far from the goal of this paper and, yet, one cannot help but notice the array of leading figures of this period who made an appearance in this account of the development of standard measurement units.  It is for that reason that we note the absence of Michael Faraday (1791-1867) from the above narration.  Faraday's contributions were broad, deep and central to the understanding of electromagnetism.   However, his work, such as the discovery of electromagnetic induction, does not enter directly into the definition of standard units.  Hence his absence from the discussion (other than the farad, the unit of capacitance named in his honor and listed in Table~\ref{conversion}).

Gaussian units play a role in several areas of contemporary physics.  Although Gauss initiated both the use of metric units in studies of terrestrial magnetism, and a form of the emu units, he was not the originator of Gaussian units.  Gaussian units were introduced in the 1880s by Herman von Helmholtz (1821--1894) and Heinrich Hertz (1857--1894).  Hertz is best known for a series of remarkable experiments in which he  demonstrated the truth of Maxwell's electromagnetic theory by propagating electromagnetic waves and measuring their properties.  In the course of his work he simplified the form of Maxwell's equations (in parallel with Heaviside) and also arranged the equations symmetrically with regard to the appearance of factors of $c$.\cite{hertz}   Hertz's formulation was influenced by an 1882 paper\cite{helm}  in which Helmholtz named a system of units in honor of Gauss. 

Hendrik Lorentz (1853--1928), a leading physicist at the turn of the century, favored the Gaussian system of units, but he also favored Heaviside's idea of rationalization.  The rationalized Gaussian system is referred to as the Heaviside-Lorentz system.  We have now identified five different systems of units.\cite{desl}  Reference~\onlinecite{jackson}  provides a clear analysis of what the possibilities and constraints are with regard to defining a consistent set of units. 

The architects of the SI system of units chose to define current, with dimensions of coulombs/second, as the fourth fundamental unit rather than the coulomb itself, because current was seen as more amenable to absolute measurement.  Before current was selected, resistance had been a prime contender as the fourth basic unit because of the relative ease of representing it  by physical secondary standards.  The number or identification of the fundamental units does not affect the physics.\cite{birge} 

Finally, we reflect upon a phrase in the abstract, “``a theretofore undefined MKS system'', which referred to Giorgi's 1901 proposal.  Could that really have been the case, given that the French artifacts of 1799 were the meter and the kilogram?  Gauss and Weber had worked with millimeters and milligrams.  No doubt, in some field, in some publication, MKS units had been used, but with regard to electric and magnetic units it appears not, and it was through these units that all five of the modern systems arose.    The B.A.A.S. committee in their very first report, made clear that they were seeking a self-consistent set of practical electrical and magnetic units and that these units should bear a definite relation to the unit of work.  We have to believe that in the forty years between that report and Giorgi's proposal, many people had recognized the equality of the joule with the expression of ergs in MKS units.  Yet, they were wed to the concept of an absolute system with three basic units.  It took a bold young engineer at the dawning of a new century to marry a happy coincidence to a fourth basic unit, and thus set the stage for the International System of Units.

\begin{acknowledgments}

The author thanks William D. Friedman for reading the manuscript and for providing helpful comments.  He further thanks one of the reviewers for several discerning suggestions.

\end{acknowledgments}


\begin{thebibliography}{5}
\bibitem{bord2} Ref.~\onlinecite{bord}, history of SI units can be found in concluding chapter.
\end{thebibliography}
\end{document}